\pdfoutput=1
\documentclass[12pt,preprint]{aastex}

\usepackage{graphicx}

\newcommand{\fluxu}{erg s$^{-1}$ cm$^{-2}$} 

\newcommand{\ilamu}{erg s$^{-1}$ cm$^{-2}$ arcsec$^{-2}$ \AA$^{-1}$}
\newcommand{\sbu}{erg s$^{-1}$ cm$^{-2}$ arcsec$^{-2}$}

\newcommand{\kms}{km s$^{-1}$}
\newcommand{\mum}{$\mu$m}

\newcommand{\iras}{\textit{IRAS}}
\newcommand{\spitzer}{\textit{Spitzer}}

\newcommand{\pheii}{\ion{He}{2}}

\newcommand{\pcii}{\ion{C}{2}}
\newcommand{\pciv}{\ion{C}{4}}
\newcommand{\iciii}{\ion{C}{3}]}

\newcommand{\nii}{[\ion{N}{2}]}
\newcommand{\pnv}{\ion{N}{5}}
\newcommand{\iniii}{\ion{N}{3}]}
\newcommand{\iniv}{\ion{N}{4}]}

\newcommand{\oii}{[\ion{O}{2}]}
\newcommand{\oiii}{[\ion{O}{3}]}
\newcommand{\oiv}{[\ion{O}{4}]}

\newcommand{\povi}{\ion{O}{6}}
\newcommand{\ioiii}{\ion{O}{3}]}
\newcommand{\ioiv}{\ion{O}{4}]}

\newcommand{\neii}{[\ion{Ne}{2}]}
\newcommand{\neiii}{[\ion{Ne}{3}]}
\newcommand{\nev}{[\ion{Ne}{5}]}
\newcommand{\inev}{\ion{Ne}{5}]}
\newcommand{\inevi}{\ion{Ne}{6}]}

\newcommand{\silii}{[\ion{Si}{2}]}
\newcommand{\psiliv}{\ion{Si}{4}}
\newcommand{\isiliii}{\ion{Si}{3}]}

\newcommand{\sii}{[\ion{S}{2}]}
\newcommand{\siii}{[\ion{S}{3}]}
\newcommand{\siv}{[\ion{S}{4}]}

\newcommand{\ariii}{[\ion{Ar}{3}]}

\newcommand{\pcaii}{\ion{Ca}{2}}

\newcommand{\feii}{[\ion{Fe}{2}]}
\newcommand{\feiii}{[\ion{Fe}{3}]}

\slugcomment{Accepted for publication in the ApJ.}

\shorttitle{IRS spectroscopy of CygXA}
\shortauthors{Sankrit et al.}

\begin{document}

\title{\spitzer\ IRS Observations of the XA Region in the Cygnus
Loop Supernova Remnant\footnotemark[1]}

\footnotetext[1]{Based on observations made with the Spitzer Space Telescope.}

\author{Ravi Sankrit\altaffilmark{2}, 
John C. Raymond\altaffilmark{3},
Manuel Bautista\altaffilmark{4},
Terrance J. Gaetz\altaffilmark{3},
Brian J. Williams\altaffilmark{5},
William P. Blair\altaffilmark{6},
Kazimierz J. Borkowski\altaffilmark{7}, 
\and
Knox S. Long\altaffilmark{8}
}

\altaffiltext{2}{SOFIA Science Center, NASA Ames Research Center, M/S N211-3, Moffett Field, CA 94035.}
\altaffiltext{3}{Smithsonian Astrophysical Observatory}
\altaffiltext{4}{Western Michigan University}
\altaffiltext{5}{Goddard Space Flight Center}
\altaffiltext{6}{Johns Hopkins University}
\altaffiltext{7}{North Carolina State University}
\altaffiltext{8}{Space Telescope Science Institute}

\begin{abstract}

We report on spectra of two positions in the XA region of the Cygnus
Loop supernova remnant obtained with the InfraRed Spectrograph on
the Spitzer Space Telescope.  The spectra span the 10--35\,\mum\
wavelength range, which contains a number of collisionally excited
forbidden lines.  These data are supplemented by optical spectra
obtained at the Whipple Observatory and an archival UV spectrum
from the International Ultraviolet Explorer.  Coverage from the UV
through the IR provides tests of shock wave models and tight
constraints on model parameters.  Only lines from high ionization
species are detected in the spectrum of a filament on the edge of
the remnant.  The filament traces a 180 \kms\ shock that has just
begun to cool, and the oxygen to neon abundance ratio lies in the
normal range found for Galactic H~II regions.  Lines from both high
and low ionization species are detected in the spectrum of the cusp
of a shock-cloud interaction, which lies within the remnant boundary.
The spectrum of the cusp region is matched by a shock of about 150
\kms\ that has cooled and begun to recombine.  The post-shock region
has a swept-up column density of about $1.3\times 10^{18}$~cm$^{-2}$,
and the gas has reached a temperature of 7000 to 8000~K\@.  The
spectrum of the Cusp indicates that roughly half of the refractory
silicon and iron atoms have been liberated from the grains.  Dust
emission is not detected at either position.

\end{abstract}



\section{Introduction}

Supernova remnants (SNRs) play an important role in the life-cycle
of dust in the interstellar medium (ISM).  As SNR shock waves sweep
up interstellar material they heat the gas and dust, and they destroy
a significant fraction of the grains, whereby refractory elements
are released back into the gas phase.  The shock-heated dust emits
strongly at infrared (IR) wavelengths and is the major contributor
to the total IR flux from remnants \citep{arendt89, saken92}.  The
IR wavelength regime also contains a number of collisionally excited
lines that are emitted by radiative shocks.  These lines provide
diagnostics for the gas-phase elemental abundances, and a comparison
of refractory and non-refractory species can yield a measurement
of the efficiency of grain destruction.  Thus, IR observations of
SNR shocks are crucial for studying the dust destruction process
in shocks, and useful for studying the shock properties.  Most SNRs
in the Galaxy are highly extincted and they cannot be detected at
ultraviolet or even optical wavelengths, and in those cases the IR
emission offers the only way to study the radiative shocks.

The Cygnus Loop, a middle-aged remnant, is an ideal object for the 
study of SNR shocks.  It is bright and it is nearby, so the emitting regions 
can be studied at high spatial resolution.  It is located away from the 
Galactic mid-plane and the foreground extinction is low, and is therefore
observable in the ultraviolet and far-ultraviolet wavelength regimes.

The Cygnus Loop exhibits a classical shell morphology at all
wavelengths.  In the IR, this is clearly seen in images obtained
by the \textit{Infrared Astronomical Satellite (IRAS)} \citep{braun86}.
\citet{arendt92} carried out an analysis of the \iras\ data, where
they decomposed the IR emission into two components, one associated
with the X-ray limb of the remnant and the other with the bright
optical regions.  The two components correspond to non-radiative
and radiative shocks, respectively.  \citet{arendt92} concluded
that the component associated with the X-ray limb was due to emission
from thermal dust.  For the component associated with the optical
regions, they estimated that between 10\% and 100\% of the emission
in the broad band \iras\ images could be due to IR line emission,
as opposed to dust continuum emission.

These bright optical regions in the Cygnus Loop provide the opportunity
to study the IR emission lines of a radiative shock running into
atomic gas.  Most IR spectra of SNRs to date pertain to dust emission
\citep{sankrit10, winkler13}, shocks in dense molecular clouds
\citep{oliva99a, oliva99b, neufeld07, hewitt09}, or shocks in SN
ejecta \citep{ghavamian09, rho09, temim06, temim10}.  By observing
the interaction regions in the Cygnus Loop, we can study the speeds,
compositions and swept-up column densities of moderate speed (100--200
\kms\ ) shock waves in interstellar regions.

In this paper, we focus on spectra of the well-studied ``XA'' region,
obtained with the \textit{Spitzer Space Telescope (Spitzer)}.  The
XA region is an indentation in the X-ray shell along the southeast
perimeter of the Cygnus Loop.  It was so named by \citet{hester86}
who showed that it was an interaction region between the blast wave
and a large cloud.  \citet{szentgyorgyi00} obtained narrowband
\nev\,$\lambda$3426 images of the Cygnus Loop and in the XA region
they identified a ``boundary shock'' - a long \nev\ filament with
very little associated H$\alpha$ emission.  \citet{danforth01}
analyzed optical and ultraviolet data of the XA region and suggested
that it was a protrusion on the surface of a much larger cloud.  In
their picture, the boundary shock is the one traveling through the
cavity wall, while a slower shock is being driven into the tip of
the finger, and results in bright optical and X-ray emission.  Based
on spectra obtained with the \textit{Far Ultraviolet Spectroscopic
Explorer (FUSE)}, \citet{sankrit07} showed that shocks with velocities
spanning the range 120--200\,\kms\ are present in the XA region and
that they are effective at liberating silicon from grains.  They
also showed that the boundary shock has a velocity of $\sim 180$\,\kms.

We present \spitzer\ observations and supplementary ground-based
optical spectroscopy obtained at the Whipple Observatory in \S2.
The IR results are presented in \S3, and the optical results in
\S4.  The analysis and discussion are presented in \S5, and our
conclusions are given in \S6.


\section{Observations}

Spectra of the Cygnus Loop XA region were obtained with the Infrared
Spectrograph (IRS) on \spitzer\ \citep{houck04} in 2006 June as
part of the Cycle 2 Guest Observer program 20743.  The observations
were taken in ``stare'' mode using the Short-High (SH) and Long-High
(LH) modules.  The exposure times at each location were about 480\,s
for the SH observations and 240\,s for the LH observations.
Table~\ref{tblobs} provides a summary of the IRS observations
presented in this paper.

A 24\,\mum\ image using the Multiband Imaging Photometer for \spitzer\
(MIPS) was obtained as part of the program.  The XA region was
observed for a total exposure time of about 720\,s in December 2005.
The fully reduced post-BCD data, downloaded from the \spitzer\
archive were of excellent quality, and are used in this paper.

\begin{figure}
\centering
\includegraphics[height=7.0in]{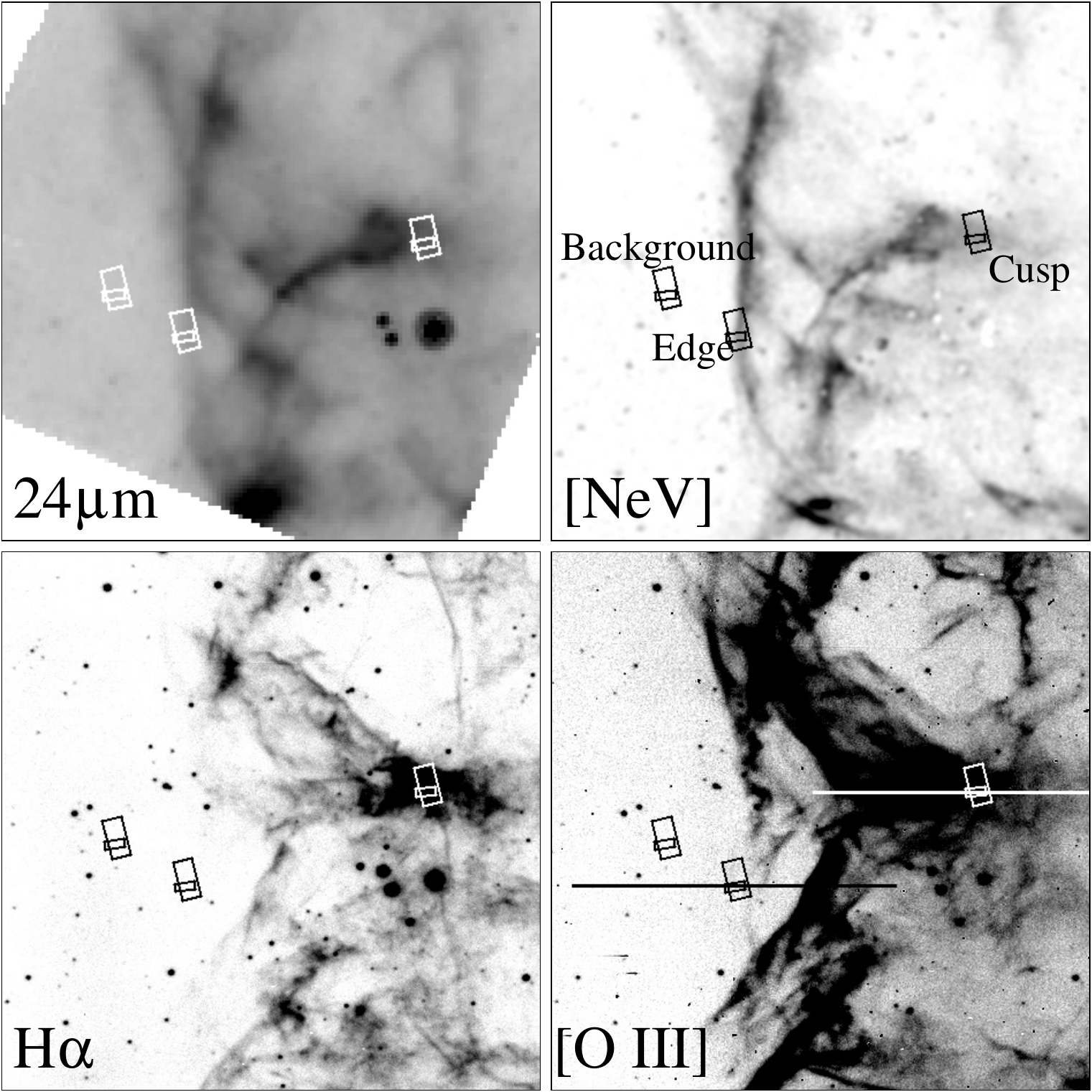}
\caption{Images of the XA region, with IRS apertures overlaid.
\textit{Left, top:} Spitzer MIPS 24\mum\ image, \textit{Right, top:}
\nev\,$\lambda$3426 image, \textit{Left, bottom:} H$\alpha$ image,
and \textit{Right, bottom:} \oiii\ image, stretched to show the
faint emission.  The boxes represent IRS apertures drawn to scale,
Short-High $4.7\arcsec\times11.3\arcsec$ and Long-High
$11.1\arcsec\times22.3\arcsec$.  The positions are labeled on the
\nev\ image.  The locations where long-slit spectra were obtained
are shown on the \oiii\ image.  (Note that the Western edge of the
slit at the Cusp position extended beyond the image field of view.)}
\label{fimages}
\end{figure}

The MIPS 24\,\mum\ image of the XA region is shown in the top-left
panel of Figure~\ref{fimages}.  Going clockwise, the remaining
panels show \nev\,$\lambda$3426, \oiii\,$\lambda$5007, and H$\alpha$
narrowband images obtained at the 1.2\,m telescope at the Whipple
Observatory, located on Mount Hopkins, Arizona.  Overlaid on the
images are boxes showing the locations of the IRS apertures.  The
smaller boxes represent the $4.7\arcsec \times 11.3\arcsec$ SH
aperture and the larger boxes the $11.1\arcsec \times 22.3\arcsec$
LH aperture.  The labels we use for the three positions, ``Background'',
``Edge'', and ``Cusp'' are shown on the \nev\ image.  Overlaid on
the \oiii\ image are the locations of the slit positions through
which optical spectra were obtained (in September 2012) using the
1.5\,m Tillinghast telescope, also at the Whipple Observatory.

For the spectroscopic observations, we obtained the Basic Calibrated
Data (BCD) and related ``Level 1'' files (AOR key 15087872) from
the \spitzer\ archive.  We used  the ``Spectroscopic Modeling
Analysis and Reduction Tool'' (SMART) version 8.1.2 \citep{higdon04,
lebouteiller10} to extract background-subtracted, one-dimensional,
calibrated data.  The observations of the Background region outside
the remnant were used for background subtraction.  In order to check
for consistency, we used the Spitzer IRS Custom Extraction (SPICE)
software to extract the spectra from the Cusp region.  The extraction
was performed on only one of the two nod positions.  The background
was subtracted, but there was no interactive removal of bad pixels.
Due to the remaining bad pixels, the resulting spectra were noisier
than the ones extracted using SMART, but the line positions and
fluxes in the two spectra matched extremely well.  In this paper,
we use only the spectra that were extracted using SMART.

The long-slit optical spectra were obtained using the FAST spectrograph
\citep{fabricant98} through a 2\arcsec\ wide slit.  The slit length
is about 3\arcmin\ and the spatial axis was oriented in the E-W
direction, and centered close to the Short-High aperture positions.
The spectrophotometric standards BD+253941, BD+174708 and HD19445
were also observed, and used to calibrate the XA region data.  The
2-D spectra have an angular scale of 0.57\arcsec/pixel and are
binned by 2 pixels, and they have a wavelength scale of 1.472\AA/pixel.

For the Cusp region, we also include an \textit{International
Ultraviolet Explorer} (IUE) spectrum reported by \citet{danforth01}.
Their position 1M coincides with the Spitzer SH position, although
the 10\arcsec\ by 20\arcsec\ IUE aperture is much larger.  We discuss
the scaling of the fluxes in the different apertures below.


\section{Results from the Infrared Spectra}

The two positions selected for observation sample different shock
interactions.  The Edge is the shock at the outer perimeter of the
remnant.  The \nev\ emission is strong, but the post-shock material
has not yet recombined and therefore the H$\alpha$ emission is
undetected (Figure\,\ref{fimages}).  The Cusp is a region of complex
morphology at the tip of a cloud protrusion, and both in \nev\ and
in H$\alpha$ emission line images.

\begin{figure}
\centering
\includegraphics[width=7.0in]{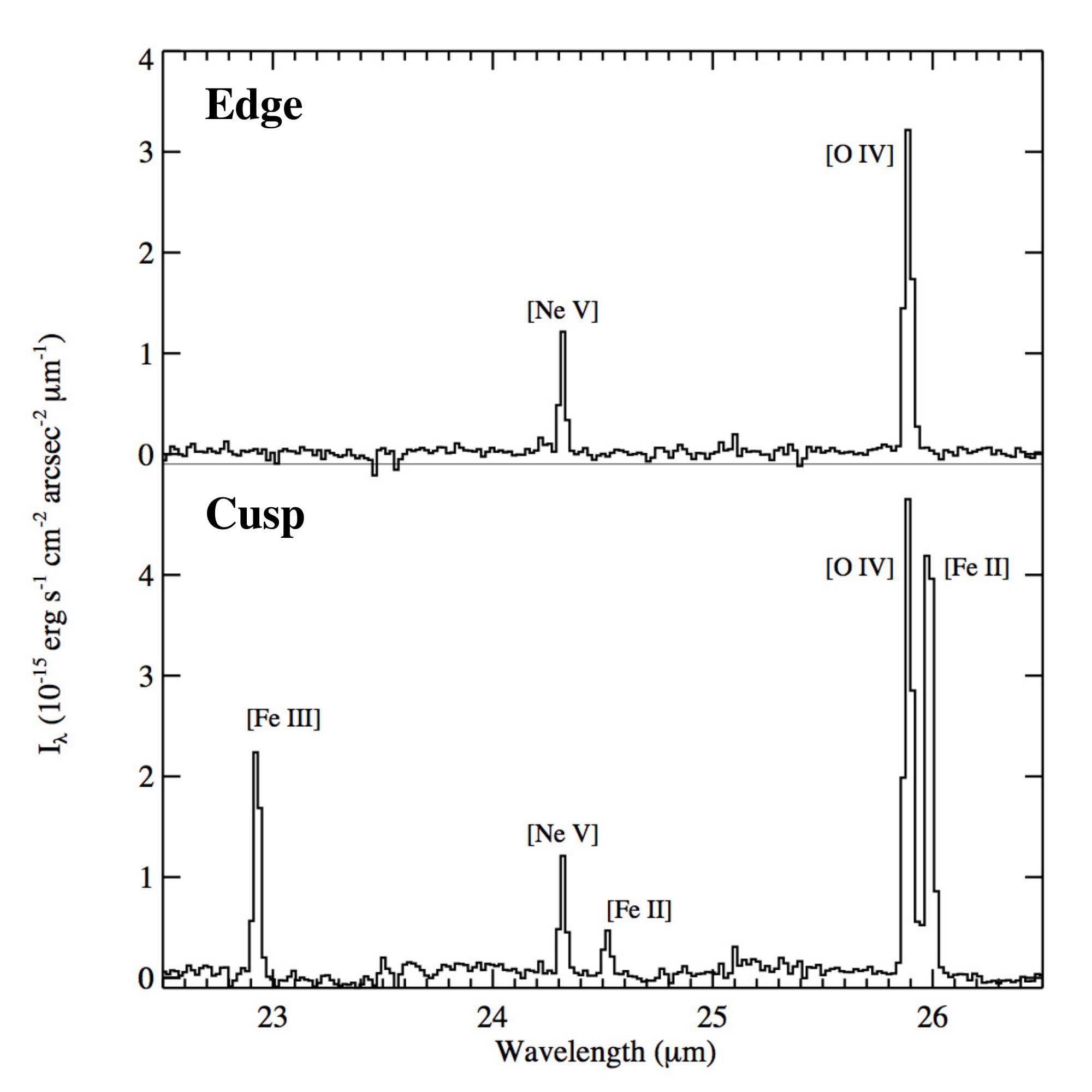}
\caption{Background subtracted Long-High spectra of the Edge and
Cusp shocks in the 22.5--26.5\mum\ wavelength region.  Note that
\oiv\ and \feii\ lines are well resolved.}
\label{fspec2327}
\end{figure}

\begin{figure}
\centering
\includegraphics[height=7.0in]{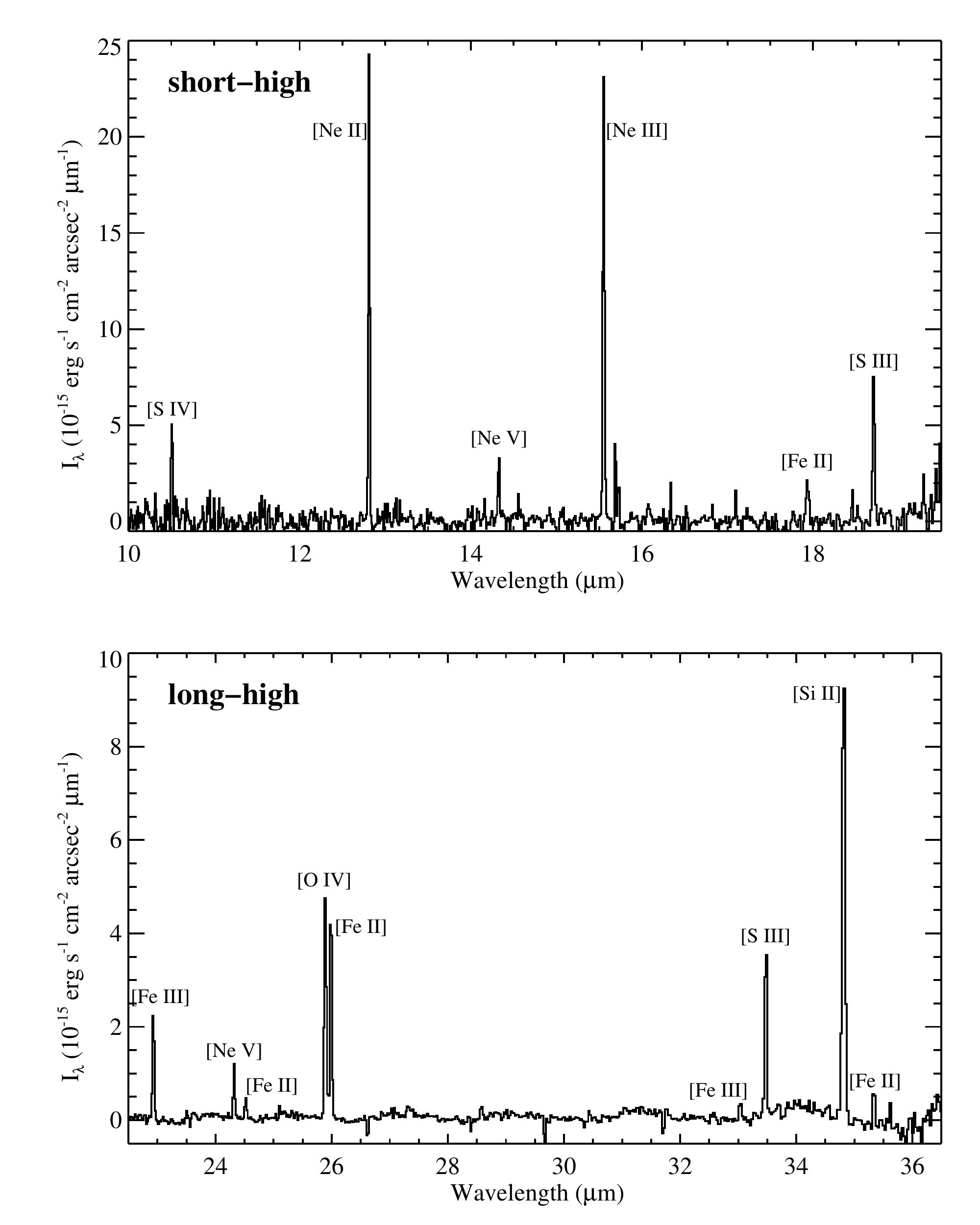}
\caption{IRS short-high (upper panel) and long-high (lower panel)
spectra of the Cusp.  The region between 19.5\,\mum\ and 22.5\,\mum\
contains no emission lines, and is not shown.  Most of the low level
spikes -- positive and negative -- are artefacts.}
\label{fcuspspecir}
\end{figure}

The Edge spectrum contains only \nev\ and \oiv\ lines, both of which
are high ionization species.  By contrast, the Cusp spectrum is
rich in lines, and includes emission from high and low ionization
species, such as \oiv, \neii, \silii\ and \feii.  Spectra in the
22.5--26.5\,\mum\ range are plotted for both locations in
Figure~\ref{fspec2327}, and the IR spectrum of the Cusp is shown
in Figure~\ref{fcuspspecir}.

The line strengths in each of the spectra were measured using both
a simple integration under the curve and gaussian fitting; the two
measurements are consistent to within the statistical errors, which
we estimate to be about 5\% for the strong lines and about 15\% for
the weak lines.  In the Edge SH spectrum the intensity of
\nev\,14.33\,\mum\ is $3.6\times 10^{-17}$\,\sbu.  In the LH spectrum,
the intensity of \nev\,24.32\,\mum\ is $4.1\times 10^{-17}$\,\sbu,
and of \oiv\,25.88\,\mum\ is $14.0\times 10^{-17}$\,\sbu.

The line intensities measured in the Cusp spectra are presented in
Table~\ref{tblflux} along with the UV line intensities from
\cite{danforth01} and optical line fluxes from FAST.  The values
reported are from the gaussian fits.

The high resolution modes of the IRS are not optimal for detecting
low-level continuum emission.  In order to check for a possible
contribution from the dust continuum, we compared the MIPS 24\,\mum\
image flux from each of the LH apertures (after subtracting the
flux from the background aperture) with the fluxes from the IRS
spectra convolved with the MIPS 24\,\mum\ filter response curve.
We find that for both the Edge and the Cusp, the IRS value was about
80\% of the MIPS value.  The emission from the XA region is completely
dominated by the collisionally excited lines, with at most 20\%
contribution from thermal dust.


\section{Results from Optical Spectroscopy}


\subsection{\oiii\ Emission from the Edge}

\begin{figure}
\centering
\includegraphics[height=7.0in]{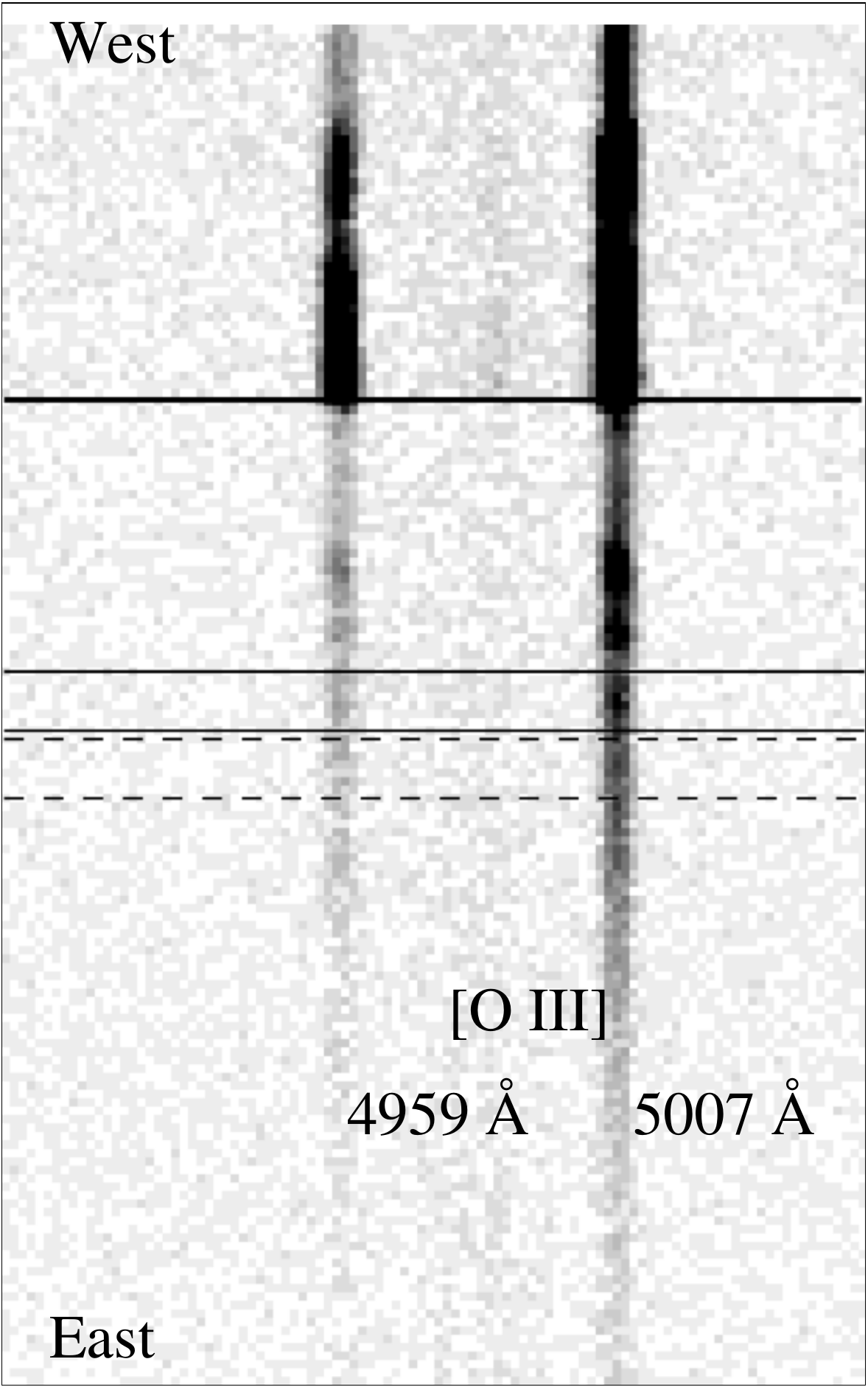}
\caption{Two-dimensional long-slit spectrum across the Edge
position showing the wavelength region around the
\oiii\,$\lambda\lambda$4959,5007 doublet.  The region corresponding
to the Edge shock is shown between the two solid lines in the middle
of the image.  The region chosen for the background lies between
the dashed lines.  The thicker solid line indicates a position about
40\arcsec\ towards the interior of the remnant demarcated by bright
optical emission, evident in the \oiii\ image (Fig.~\protect\ref{fimages}).}
\label{fedge2d}
\end{figure}

Fig.~\ref{fedge2d} shows the 2-D spectrum of the slit lying across
the Edge shock, in the wavelength region 4900--5050\AA, which
includes the \oiii\ doublet.  The spatial axis is vertical, with
East at the bottom and West at the top.  The location of the Edge
shock is between the pair of thin solid lines across the center of
the image.  The thick solid line about 3/4 the way up demarcates
the boundary of the  bright radiative filament, about 40\arcsec\
West of the Edge shock.  Faint \oiii\ emission extends beyond the
location of the Edge shock.  A few features corresponding to places
where the slit crosses shock fronts are clearly seen.  The Edge
shock is easily identified, both because it is in the middle of the
slit, and from its distance to the bright radiative shock, which
is consistent with the separation seen in the \oiii\ image
(Fig.~\ref{fimages}).

\begin{figure}
\centering
\includegraphics[height=7.0in]{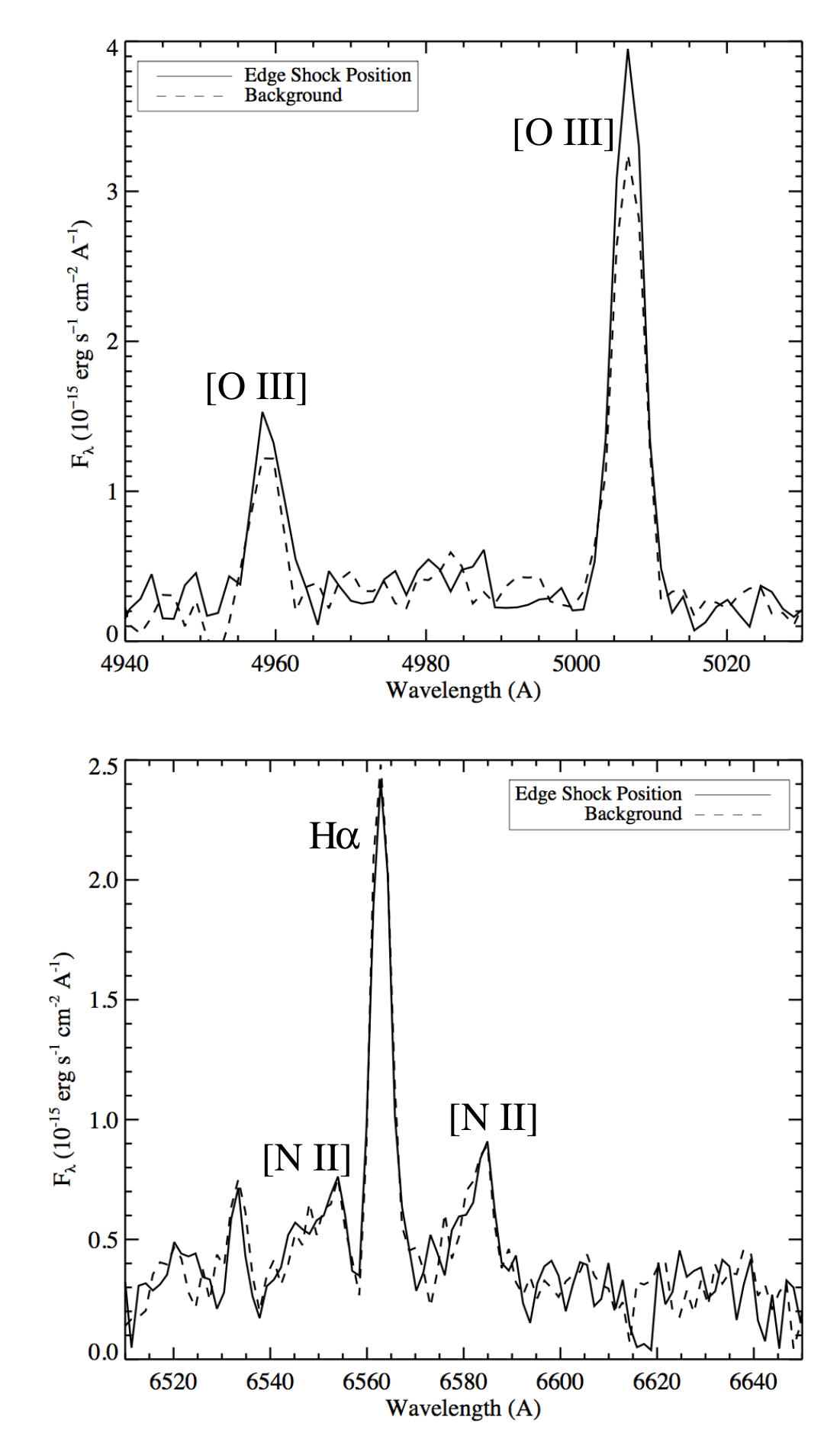}
\caption{Flux-calibrated spectra of the Edge shock and of the
background extracted from the regions shown in the 2D spectrum in
Fig.~\protect\ref{fedge2d}.  The top plot shows the \oiii\ lines,
and the bottom plot shows the wavelength region around the H$\alpha$
and \nii\ lines.}
\label{fedgespec}
\end{figure}

Spectra were extracted over 8 spatial pixels (9.12\arcsec) corresponding
to the Edge shock position, and an adjacent region of the same width
lying beyond the edge.  These are shown by pairs of solid and dashed
lines, respectively, on the image of the 2-D spectrum.  The extracted,
flux-calibrated spectra are shown in Fig.~\ref{fedgespec}.  The top
plot shows the region around the \oiii\ lines, and the excess flux
in the Edge shock position is clearly seen.  The H$\alpha$ and \nii\
are shown in the bottom plot - they are detected in both Edge and
background spectra and of equal strengths.  (The \sii\ lines, not
shown, are likewise detected at both positions with equal strengths.)
We therefore ascribe the excess \oiii\ flux in the Edge shock
position to emission from the shock itself, and the remaining as
part of the background.

We obtained the fluxes in the \oiii\,$\lambda$5007 line in each of
the spectra by fitting gaussians, and by integrating under the lines
and subtracting the backgrounds.  The methods yielded results within
3\% of each other.  We also found that the strength of the weaker
line, \oiii\,$\lambda$4959 was 1/3 of the stronger line as expected.

The \oiii\,$\lambda$5007 fluxes are $1.84\times 10^{-14}$~\fluxu\
for the Edge position and $1.52\times 10^{-14}$~\fluxu\ for the
background.  The flux due to the Edge shock, which we take to be
the difference between the two, is $3.2\times 10^{-15}$~\fluxu\ .
Assuming that the emission fills the 2\arcsec\ wide slit, and noting
the angular region over which the spectrum was extracted, the
observed \oiii\,$\lambda$5007 surface brightness of the Edge shock
is $1.75\times 10^{-16}$~\sbu\ .


\subsection{Optical Spectrum of the Cusp}

\begin{figure}
\centering
\includegraphics[width=5.0in]{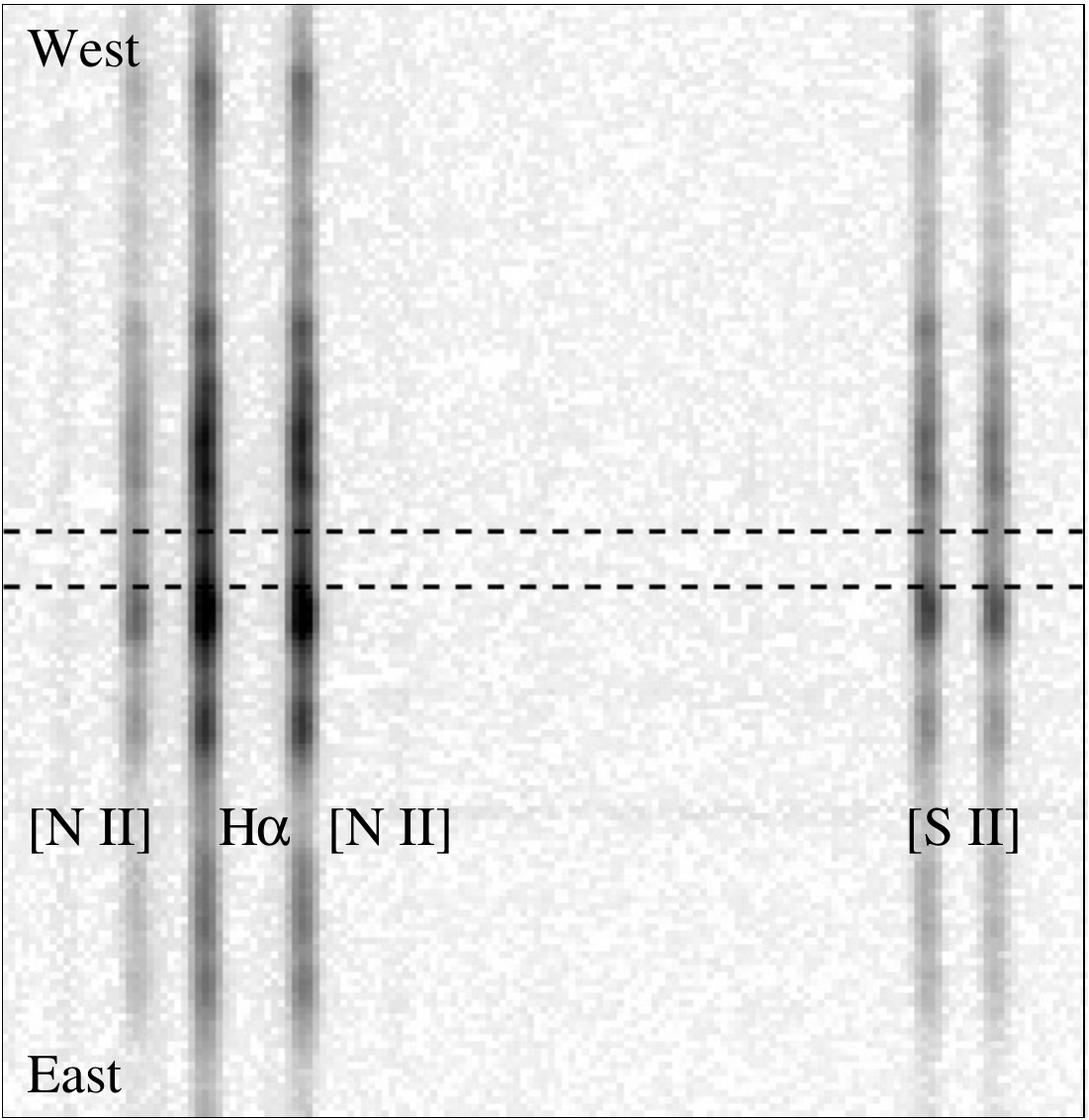}
\caption{Two-dimensional long-slit spectrum across the Cusp
position showing the wavelength region between 6520\,\AA\ and
6750\,\AA\@.  The emission lines, \nii\,$\lambda\lambda$6548,6584,
H$\alpha$\,$\lambda$6563 and \sii\,$\lambda\lambda$6716,6731 are
visible along the entire slit.  The dashed lines demarcate the Cusp
region targeted in the \spitzer\ spectra.}
\label{fcusp2d}
\end{figure}

\begin{figure}
\centering
\includegraphics[width=7.0in]{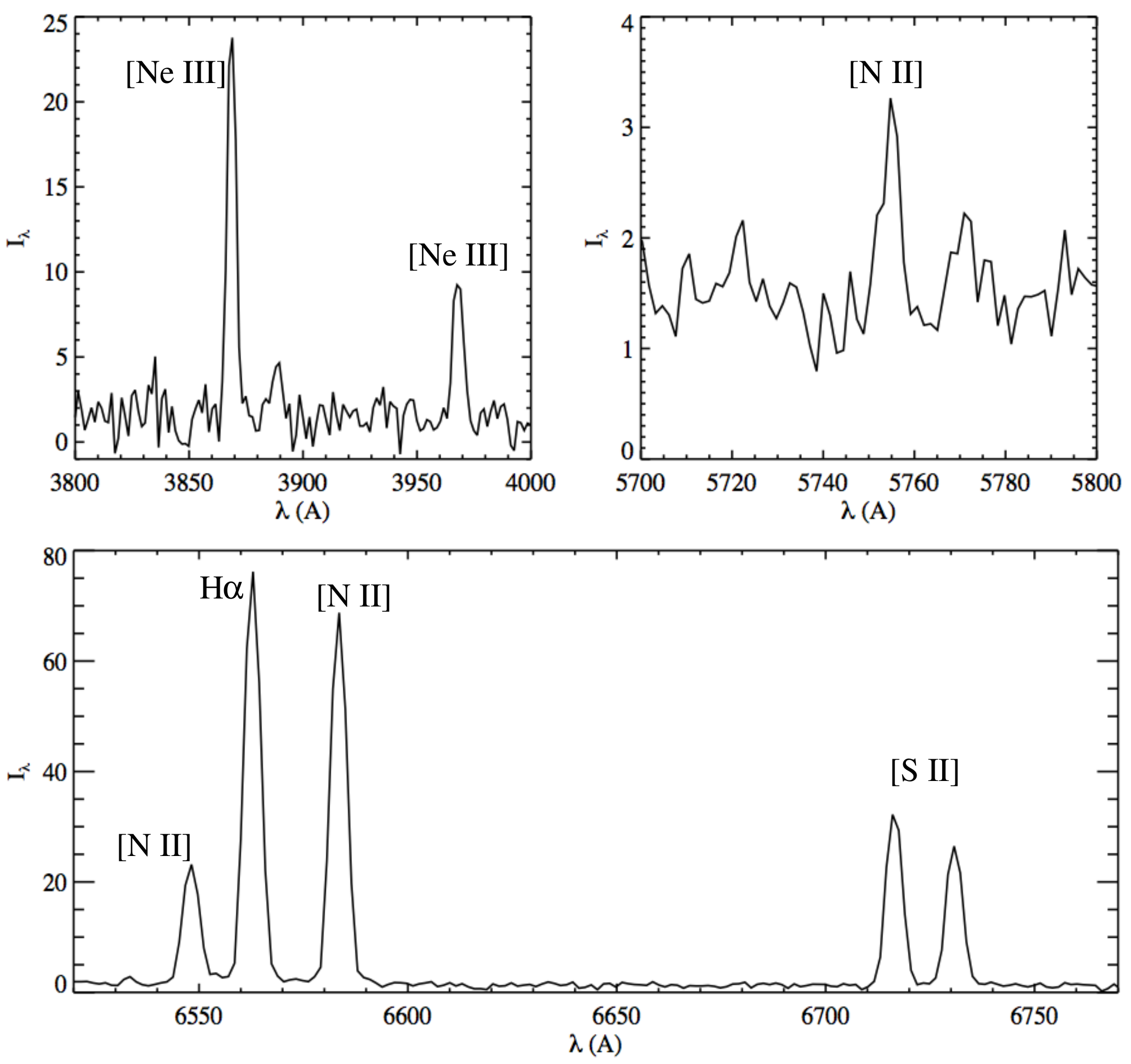}
\caption{Selected wavelength regions of the Cusp spectrum, which
was extracted along the region between the dashed lines shown in
Fig.~\protect\ref{fcusp2d}.  The intensity units are $10^{-17}$\,\ilamu.}
\label{fcuspspec}
\end{figure}

Fig.~\ref{fcusp2d} shows the 2-D spectrum of the Cusp between
6520\AA\ and 6750\AA, which includes \nii, H$\alpha$ and \sii\ lines
(labeled on the image).  The spatial axis is vertical with East at
the bottom and West at the top.  The region between the dashed lines
in the figure is approximately that overlapping the IRS apertures.
We extracted the 1-D spectrum from that region, 10.26\arcsec\ wide,
and measured the line fluxes, which are reported in Table~\ref{tblflux}.
Selected regions of the spectrum, illustrating the range of line
strengths measured, are plotted in Fig.~\ref{fcuspspec}.  The
statistical errors are $\leq 2$\% for fluxes $\gtrsim 100\times
10^{-17}$~\sbu.

The Balmer decrement and the \oii\ I$_{7330}$/I$_{3727}$ ratio from
the optical spectrum of the Cusp indicate that E$_{B-V}\sim 0.16$.
The usual reddening assumed for the Cygnus Loop is E$_{B-V}=0.08$,
but \citet{fesen82} have measured a range of Balmer decrements in
optical spectra, and they suggest an additional 0.05--0.10 magnitudes
of reddening towards certain filaments.  In Table~\ref{tblflux} we
present the reddening corrected fluxes for E$_{B-V}=0.16$.

In the Cusp spectrum, the ratio between the components of the \sii\
doublet, F$_{6716}$/F$_{6731} = 1.24\pm0.03$.  For temperatures in
the range 5000--20,000\,K, this implies electron densities in the
range 100--250\,cm$^{-3}$.  (The calculations reported in this
section were done using PyNeb\footnote{available at
http://www.iac.es/proyecto/PyNeb/} \citet{luridana12}.)

\nii\,$\lambda$5755 is detected in our Cusp spectrum at a level
$\geq 3\sigma$ (Fig.~\ref{fcuspspec}).  The temperature sensitive
ratio I$_{[5755]}$/I$_{[6548+6584]}$ equals $(10\pm3)/429$
(Table~\ref{tblflux}), which implies that the temperature is about
$14,200\pm\, 2300$ K, consistent with the values used in the density
calculation, above.  The temperature-sensitive ratio \oiii\
I$_{[4363]}$/I$_{[5007+4949]}$ is 0.08, which indicates that the
\oiii\ emission arises at about 80,000 K, a much higher temperature
than the typical average value of 25,000 K found in a complete
radiative shock.  That implies that the shock is incomplete, so
that the swept-up column density is small enough that radiative
cooling \citep{raymond88} has only brought the temperature the gas
shocked earliest to 7000 to 8000 K, and recombination is far from
complete.  This provides a constraint on the models, discussed
below.

Fig.~\ref{fcusp2d} shows that emission is detected across the entire
slit, and variations in the line intensities are evident.  In order
to check how sensitive our results were to the exact region of
extraction, we compared the Cusp spectrum with the spectrum of the
brighter knot lying adjacent and to the East of the Cusp.  The lines
are brighter by a factor of $\sim$1.4 but the relative fluxes are
approximately the same.  In particular, the \sii, \nii\ and \oiii\
density and temperature sensitive line-ratios are consistent within
the error bars.  Also, F$_{{\rm{[S\,II]}}}$/F$_{{\rm{H}}\alpha}
\approx 0.8$ in both spectra.


\section{Analysis and Discussion}

For non-uniform extended emission observations with IRS the relative
calibration of the SH and LH modules is highly uncertain (L. Armus,
private communication).  The absence of any continuum emission means
that spectra from the two modules cannot be easily normalized.
Additionally, since the SH aperture covers only about one-fifth the
area of the LH aperture, the two sample different emitting regions.
However, the ratio of the \nev\ lines at 14.33$\mu$m and 24.32$\mu$m
is determined solely by the excitation rates in the density and
temperature range of interest, and we use that ratio to scale the
fluxes (see below).

The shock models were calculated using the code originally described
by \citet{raymond79} with updates described in \citet{raymond97}.
We have further updated the code to use the recombination rates
used by the CHIANTI package version 7.1 \citep{dere97, landi13},
which corrects errors in the rates for singly and doubly ionized
species in version 6.  We have also updated the collision strengths
for the IR lines, many of which have strong contributions from
resonances.  The data used include \citet{aggarwal08} for \oiv,
\citet{witthoeft07} for \neii, \citet{mclaughlin11} for \neiii,
\citet{dance13} for \nev, \citet{bautista09} for \silii, \citet{grieve13}
for \siii\ and \citet{tayal00} for \siv.  In addition, at the
temperatures where \ion{O}{4} and \ion{Ne}{5} are found in collisionally
ionized plasmas, the contributions of excitation by proton collisions
and of cascades from excitations to higher levels can be important.
We used CHIANTI to include those effects.  The \feii\ line intensities
were computed from the 52-level model of \citet{bautista98}, under
the assumption of pure collisional excitation, and the \feiii\ line
intensities were calculated using the 36-level model of \citet{bautista10}.


\subsection{The Edge Shock}

A far-ultraviolet spectrum of the Edge shock, obtained using
\textit{FUSE}, was analyzed by \citet{sankrit07}.  They found that
the spectrum could be produced by a 180\,\kms\ shock and a swept-up
column density $1.66\times 10^{18}$~cm$^{-2}$.  At this stage of
shock-completeness (swept-up column density at a given shock speed),
the Ne$^{4+}$ zone is nearly complete, while the O$^{3+}$ zone is
about 85\% complete, and O$^{2+}$ zone is just starting to develop.
Thus the \oiii\,$\lambda$5007 to \oiv\ ratio, which is independent
of the oxygen abundance, is very sensitive to the swept-up column,
while the \oiv\ to \nev\ flux ratio is less sensitive.

From \S4.1 and \S3, the observed \oiii\ to \oiv\ flux ratio is 1.25.
Corrected for interstellar extinction, using E$_{B-V}$=0.08 towards
the Cygnus Loop \citep{parker67}, R$_{V}=3.1$, and the prescription
of \citet{cardelli89} implemented in the IDL astronomy library of
routines, the ratio is $\approx 1.6$.  (Note that we have used
E$_{B-V}$=0.16 for the Cusp region -- this difference will be
discussed further in \S5.2, below.) The Edge shock does not fill
the Long-High aperture, as is evident from the \nev\ image shown
in Fig.~\ref{fimages}, and so the reported surface brightness is
likely to be an underestimate.  If we assume that the filling
fraction is 0.5, then the resulting \oiii\ to \oiv\ would be $\approx
0.8$.

\begin{figure}
\centering
\includegraphics[width=7.0in]{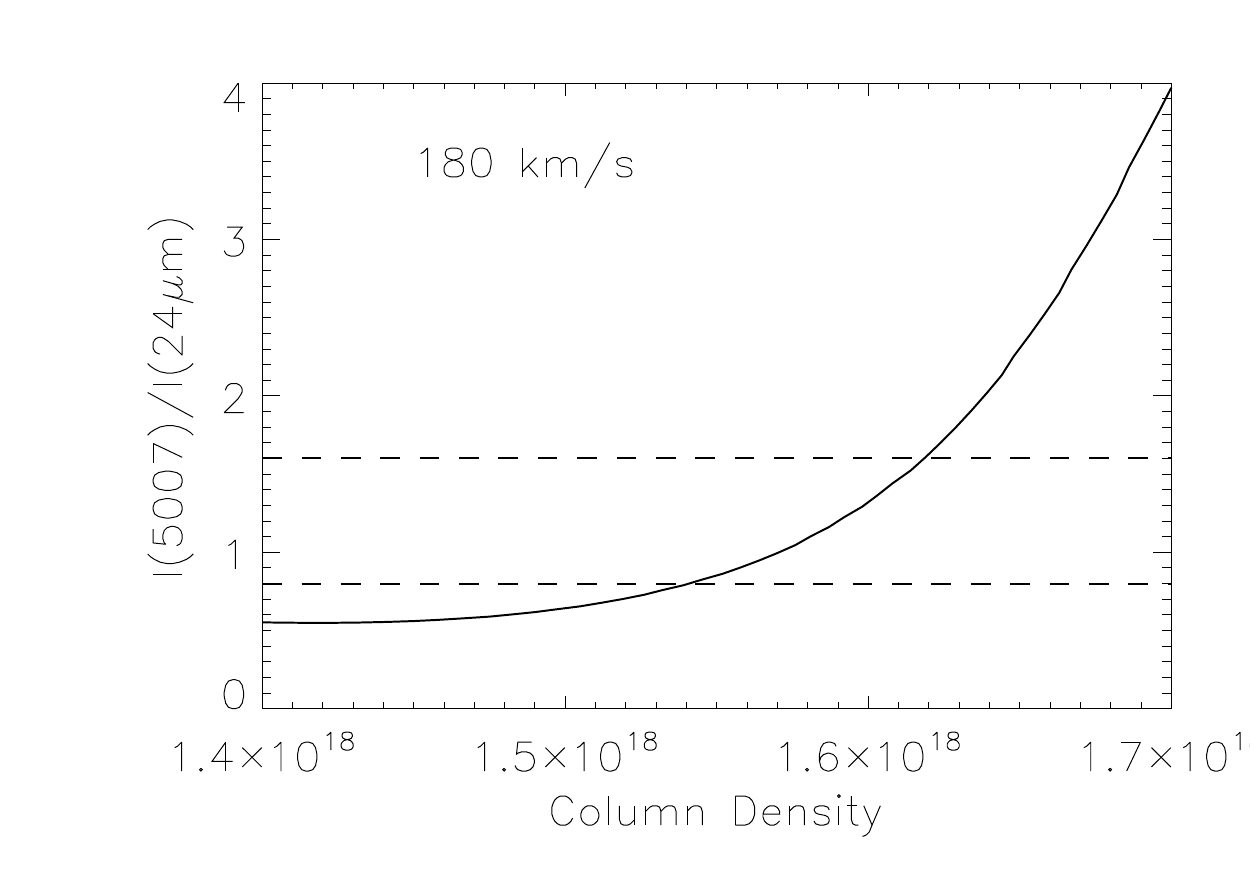}
\caption{The plot shows the flux ratio between \oiii\,$\lambda$5007
and \oiv\,25.88\mum\ as a function of swept-up column density for
the fiducial model.  The dashed lines indicate the points where the
prediction matches the lower and upper bounds of the observed ratio.
See \S5.1 for details.}
\label{fo3too4}
\end{figure}

In Fig.~\ref{fo3too4} the \oiii\ to \oiv\ flux ratio is plotted as
a function of swept-up column for a 180 \kms\ shock.  The dashed
lines mark the two points bracketing the observed value -- 0.8 and
1.6.  At these points, for O=8.70 and Ne=8.09, the values of the
\oiv\ to \nev\ flux ratio are 1.6 and 2.2, respectively.  For a
given shock velocity, and for a particular swept-up column this
flux ratio is directly proportional to the abundance ratio, O/Ne.
In the model, O/Ne = 4.1, and the observed value of \oiv\ to \nev\
is 3.4.  Scaling the model to match the observed values at the two
points, we find that the required values of O/Ne are 8.9 and 6.3
for the lower and higher swept-up columns.

The \nev\ emissivity is sensitive to the shock velocity around the
180\,\kms\ region, with faster shocks producing brighter \nev\
emission.  The shock velocity is well constrained to be 180\,\kms\
by the \inevi\ to \inev\ ratio in the FUV spectrum \citep{sankrit07}.
The observed \inevi\ emission would not be produced by slower shocks.
Therefore, we performed the calculation described above only for a
faster, 190\,\kms, shock, and found that O/Ne in the range 9.1 to
13.2 are required.

Our choice of the oxygen to neon abundance ratio for the original
model was based on its success in reproducing the FUV line strengths
\citep{sankrit07}.  However, updates to the atomic rates (see \S5,
above) have resulted in a revision of the ratio of abundances.
Furthermore, the forbidden optical and IR lines are not susceptible
to uncertainties due to resonance scattering, and therefore the
results presented here are more robust than those from the earlier
study.  The abundance ratio O/Ne=6.3 is close to the median value
of 5.89 found for a sample of 10 Galactic H~II regions \citep{shaver83}.
The maximum value of the ratio was 8.9 in the sample.  Thus, the
O/Ne we find for the 180\,\kms\ shock are in the normal range for
H~II regions, while the high values found for the 190\,\kms\ shock
are implausible.


\subsection{The Cusp Region}

To compare the observations of the Cusp with theoretical models,
we must apply a reddening correction and scale the fluxes from the
different apertures to account for the variation in surface brightness
within the larger apertures.  The value for E$_{B-V}$ obtained from
the optical spectrum differs from the usual value for the Cygnus
Loop (\S4.2).  As a check, we extracted the flux in the \nev\,3425\,\AA\
line in the region of the \spitzer\ SH aperture from the image
presented by \citet{szentgyorgyi00} and compared it with the
\nev\,14.33\,\mum\ flux, and the ratio agrees with the theoretical
value and E$_{B-V}=0.08$.

The difference in the values of E$_{B-V}$ derived from the Balmer
decrement and the \oii\ lines on the one hand, and from the \nev\
optical and IR lines on the other is consistent with the idea that
on average the Balmer lines and low-ionization lines arise from
deeper within the interaction region and the high-ionization lines
from the outer, less extincted regions.  The reddening, E$_{B-V}=0.08$
is the normal foreground, and the excess is local.  An extended
component of high ionization emission has been observed in the XA
region.  \textit{FUSE} spectra presented by \citet{sankrit07} shows
the presence widespread \povi\,$\lambda\lambda$1032,1038 emission,
and strong \povi\ emission, comparable in strength to the
\ioiv\,$\lambda$1400 line, was observed in the area by HUT
\citep{danforth01}.  These provide some evidence for a faster shock,
or for a transition layer where cool gas evaporates to join the
X-ray emitting plasma that envelopes the optical filaments.  However,
the 56\arcsec\ HUT aperture encompasses a lot of material outside
the region observed by the other instruments considered here, so
there is no compelling evidence for a faster shock contributing
specifically to the spectra we have observed.  An alternative
explanation of the discrepancy is that we have underestimated the
error in the surface brightness measured from the narrowband image,
both due to the calibration and due to the placement of the SH
aperture FOV on the image, and the result is consistent with the
higher value of reddening.

Faced with these possibly contradictory results, we calculated the
corrected spectrum for both values of reddening, using the
\citet{cardelli89} extinction function with R$_{V}=3.1$.  After
applying the reddening correction, we accounted for the large
variations of intensity within the different apertures by scaling
the fluxes to match theoretical intensity ratios.  Ideally intensity
ratios that depend only on the atomic rates, and are therefore
insensitive to the model should be used.  The ratio of the two IR
lines of \nev\ is one such ratio.  To scale the LH aperture to the
SH we use the \nev\,14.33\,\mum\ to 24.32\,\mum\ ratio=0.82.  This
means multiplying the LH values by a factor of 1.90, which is
plausible because the average flux in the LH aperture is presumably
lower than in the SH aperture, which was placed on the intensity
peak as judged by the optical images.  This scaling greatly improves
the agreement of the \siii\,18.71\,\mum\ to 33.47\,\mum\ ratio.
(We use the \nev\ ratio rather then the \siii\ ratio, because the
latter depends on density in the range indicated by the \sii\ optical
lines.)

For the optical and UV fluxes, the only ratios available have some
temperature sensitivity, so the scaling is less certain.  To scale
the optical flux to the SH flux we use the
\neiii\,$\lambda\lambda$3870,3968 to 15.56\,\mum\ ratio of 3.87.
Since that depends on the reddening correction, the scalings differ
for the two reddenings chosen;  1.0 for E$_{B-V}=0.08$ and 0.70 for
E$_{B-V}=0.16$.  Next, we scale the UV fluxes by using the
\ioiii\,1665\AA\ to \oiii\,5007\AA\ ratio of 0.52.  That results
in multiplying the UV fluxes by 1.90 for E$_{B-V}=0.08$ and 1.0 for
E$_{B-V}=0.16$.  The reddening-corrected \nev\,$\lambda$3425
intensities were not scaled, because the flux was extracted from
the region corresponding to the SH aperture.  The uncertainty in
the placement of the aperture on the image gives a 7\% uncertainty,
and \citet{szentgyorgyi00} quote a photometric uncertainty of 30\%.
Finally, we normalized the relative fluxes to H$\beta=100$, for
comparison with the models.

The scaling for the different apertures largely cancels out the
difference in reddening correction, so that for the two values of
E$_{B-V}$, the scaled, normalized relative fluxes differed by less
than 20\% for all lines except \nev\,$\lambda$3425.  In Table~\ref{tblflux}
along with the observed fluxes, we present the reddening corrected
fluxes only for the case of E$_{B-V}=0.16$ because of its better
agreement with the model Balmer decrement.


\subsubsection{Shock Models}

The line fluxes predicted by a set of shock models are presented
in Table~\ref{tblflux}.  We chose five models to illustrate the
sensitivity to shock speed and density.  These models use the
following abundance set: H=12.00; He=10.93; C=8.52; N=7.96; O=8.70;
Ne=8.09; Mg=7.52; Si=7.60; S=7.20; Ar=6.90; Ca=6.30; Fe=7.60 and
Ni=6.30.  Hydrogen was assumed ionized and the helium singly ionized
in the pre-shock gas.  The shock velocities, pre-shock densities
and pre-shock magnetic fields used in these models are presented
in Table~\ref{tblshocks}, along with values of the swept-up column
densities and temperatures where the models were truncated.

\textbf{Shock Speed:}
The range of shock speeds 145--155\,\kms\ best reproduces the line
fluxes from the wide range of ionization states that are observed.
Shocks that are outside this velocity range produce far too little
or far too much of high ionization lines such as \pnv, \pciv, \oiv\
and \nev\ (although the permitted lines are so strongly attenuated
by resonance scattering in the shocked gas and in the ISM that they
only constrain the lower end of the range).  The specific range was
chosen to match the \pheii\,1640\,\AA\ line under the assumption
that Helium is singly ionized in the pre-shock medium.

\textbf{Density and Magnetic Field:}
The observed ratio of $1.24\pm0.03$ between the components of the
\sii\ doublet (\S4.2) implies an electron density in the emitting
gas in the range 100--250\,cm$^{-3}$.  The density in the \sii\
zone depends upon the pre-shock density and on the maximum compression
in the post-shock gas, which in turn depends on the driving pressure
of the shock and on the pre-shock magnetic field and cosmic ray
pressure. These non-thermal contributions to the pressure limit the
compression of the gas as it cools from the post-shock temperature
near 260,000 K to the \sii\ formation temperature near 10,000 K
\citep{raymond79}.  The rather low values of the pre-shock magnetic
field for most of the models were chosen in order to reach the
observed density range with a pre-shock density and ram pressure
comparable to those determined for other positions in the XA region.
For the models shown, the predicted \sii\ doublet ratio agrees with
the observed value for M145, M150 and M155a.  However, the predictions
are too low for models M155b and M155c, indicating that the density
is too high.

On the other hand, observed ratio of \feiii\,22.92\,\mum\ to
\feii\,25.98\,\mum\ is about 0.4, while the lower density models
(M145, M150 and M155a) predict ratios of about 0.1.  Even the higher
density models (M155b and M155c) only predict ratios of 0.2, while
their [S II] doublet ratios are clearly at odds with the observed
value.  The lower densities are consistent with those derived at
other shocks in the Cygnus Loop, including several positions in the
XA region, while the \feii\ ratio would suggest a density of
2000\,cm$^{-3}$, 10 to 20 times higher than found at other positions.
The easiest explanation for the discrepancy is an error in atomic
rates, but a physical explanation might be that the \feii\ arises
from a separate region, such as a slow shock that is heavily reddened
and therefore contributes little to the \sii\ lines.

\textbf{Shock Completeness:}
The models in Table~\ref{tblflux} are cut off at swept-up columns
log $N_H$ = 18.0 to 18.1.  The temperatures obtained at cut-off are
between 7000 and 8000\,K (Table~\ref{tblshocks}), which is required
by the temperature of the \oiii\ zone (\S4.2).  Although a somewhat
larger range of cutoffs is allowable, too much of a deviation would
lead to unacceptable values for line ratios such as \oii/H$\beta$
and \oiii/H$\beta$.  For the densities and magnetic fields used,
the gas has cooled for between 500 and 1000 years.  A 400\,\kms\
X-ray producing shock will travel ahead by the difference in
velocities times that age, or about 100\arcsec\ in a thousand years,
which is consistent with the position of the Edge filament.  The
observed separation is 135\arcsec, which lies within the uncertainty
in projection effects and the changing speeds of the shocks as they
pass through regions of different density.


\subsubsection{Comparing the IR spectra with Shock Models}

The parameters in the shock model have been constrained largely by
the UV and optical spectra.  If we disregard the lines strongly
affected by resonance scattering, and \nev\,$\lambda$3425, for which
the measurement and scaling is discrepant compared to the other
lines, then the shock model predictions match the observations to
within 50\%, except for \iniii\,$\lambda$1750 and \ariii\,$\lambda$7138,
and in most cases to within 25\%.  Reducing the rather
uncertain abundance of Ar by a factor of about two would bring the
\ariii\ predictions into agreement with the observed value, without
affecting any other line strengths.

In Fig.~\ref{fmodobs}, the ratio of the flux predicted by model
M150 to the observed, dereddened value is plotted against wavelength
for the lines listed in Table~\ref{tblflux}.  The top, middle and
bottom plots show the UV, optical and IR lines, respectively.  It
should be noted that no distinction is made between strong and weak
lines in the plots.  So, for instance, the match between models and
observation for the weak \feiii\ lines in the IR spectrum is actually
much better than the plot seems to imply, as can be seen by examining
Table~\ref{tblflux}.

We now turn to an examination of the IR lines, focusing on models
M145, M150 and M155a, since the other two (M155b and M155c) do not
predict the observed optical \sii\ doublet ratio.  The \nev\ lines
are most sensitive to the shock velocity, and are best matched by
the 155\,\kms\ shock.  All models predict \oiv\ fluxes that are
higher than the observation -- a factor of 1.3 for the 145\,\kms\
shock and 2.0 for the 155\,\kms\ shock.  The models reproduce the
correct \siii\ line fluxes, with the 150\,\kms\ model providing the
best match.  In contrast, the models consistently predict only about
half the observed \siv\ flux.

The \silii\ fluxes predicted by the model are significantly higher
than the observed value.  Reducing the Si abundance by a factor of
about 2.5, as expected from depletion onto grains, would bring them
into agreement.  However, the line fluxes predicted by the models
are also sensitive to the column density cut-off, since the \silii\
emission would extend far into the cooling zone.

The ratios among the \feii\ lines predicted by the models are in
agreement with the observed values, but the absolute fluxes are
higher.  As with silicon, the match between models and observation
would be improved for the \feii\ lines with Fe abundances reduced
by a factor of about two.  The models correctly predict the \feiii\
line fluxes, and therefore too low a ratio between \feiii\ and
\feii\ line strengths.  The ratio is sensitive to the photoionization
rate of Fe$^{+}$, the recombination rate of Fe$^{++}$, and especially
to the charge transfer rates, which we have taken from
\citet{neufelddalgarno87} as well as to the cutoff column density.

The models assumed solar abundances for Si and Fe, while those
elements should be almost entirely depleted onto dust grains in the
ISM.  Reductions by factors of 2.5 and 2, respectively, still imply
that about 40--50\% of these elements have been returned to the gas
phase.  The ratios of the UV lines of C III] and Si III] to the O
III] line also indicate only modest depletion.  This efficient
liberation of refractory elements from grains is a little surprising,
since only about 20--40\% of the silicon and carbon are returned
to the gas phase at a swept-up column of $2\times 10^{18}$~cm$^{-2}$
in the faster X-ray producing shock in the NE Cygnus Loop studied
by \citet{raymond13}.  The sputtering rates are greatly reduced at
the lower temperatures in 150\,\kms\ shocks, but grain-grain
collisions shatter the dust particles, producing small grains that
are more vulnerable to sputtering.  \citet{jones96} find that 10\%
to 30\% of the mass of graphite and silicates in grains is returned
to the gas phase for shock speeds below 150 \kms, so the spectra
imply a higher dust destruction efficiency.  However, the behavior
of grains in high speed collisions is uncertain.

The pair of lines \neii\ and \neiii\ present an interesting puzzle,
which may have some important consequences for our understanding
of the XA region.  Separately considered, the model predictions are
in fairly good agreement with the observations.  However, the \neiii\
to \neii\ flux ratio is always less than 1.  To match the predicted
flux ratio of about 1.3, the models would need to be truncated at
unacceptably low values of swept-up column density.  The observed
flux ratio can be explained if an additional source of photoionization
is invoked.  The extra photoionization would also be helpful in
explaining the strength of the \siv\ line and possibly the \feiii\
to \feii\ flux ratio as well.


\subsubsection{Shock Driving Pressures}

The driving pressures of the shocks M145, M150, M155a, M155b and M155c, 
n$_{0}\times v_{shock}^{2}$, are 
$\approx$\,$5.3\times 10^4$, 
$1.3\times 10^5$, 
$6.0\times 10^4$, 
$1.2\times 10^5$ and 
$1.9\times 10^5$\,cm$^{-3}$ (\kms)$^{2}$,
respectively.  These values are 3 to 10 times higher than the driving
pressure found for the Edge shock, $1.9\times 10^4$ cm$^{-3}$(\kms)$^{2}$,
by \citet{sankrit07}.  The pressure derived is proportional to the
pre-shock density, which in turn is proportional to the assumed
distance to the Cygnus Loop.  \citet{sankrit07} assumed 540~pc
\citep{blair05}, but the upper limit to the distance, 640~pc
\citep{blair09}, or an even greater value, is supported by proper
motion measurements of the non-radiative shocks in the NE Cygnus
Loop \citep{salvesen09}.  Furthermore, the reported errors on the
density allow it to be about 30\% higher.  However, even if the
driving pressure for the Edge shock is 2 times the value reported
by \citet{sankrit07}, that for the Cusp is higher by a factor between
1.5 and 5.

This may not be a problem since some pressure variations within the
XA region due to the interaction of dense clouds are likely.
We note that if the cosmic ray population in the Galaxy were
simply compressed in the shock, it could provide a substantial
fraction of the pressure support in the \sii\ emitting region for
densities (and therefore compression ratios) at the upper end of
the range allowed by the \sii\ line ratio.  However, that assumes
that the cosmic rays do not diffuse away during the cooling process.


\subsubsection{Model Limitations}

The comparison of the Cusp spectrum, from 1240\,\AA\ to 35\,\mum\
with models of radiative shocks, shown in Table~\ref{tblflux}
illustrates both the usefulness of the models and their limitations.
There are many ways in which the models can be wrong.  They assume
steady flow as the gas cools until it reaches a sharp cutoff column
density.  That assumption can be violated by thermal instability
in the cooling region \citep{innes92}, but that instability becomes
strong at speeds above 150 \kms, and it takes more than one cooling
time to develop.  It is also violated if the density, and therefore
shock speed, varies significantly over the distance in which the
shock sweeps up the observed column, in this case about 0.1 pc, and
that is very likely at the edge of a cloud.  In that case the derived
shock speed, which is determined by the highest ionization states
observed, is near the current shock speed, while lower ionization
species may have passed through a faster shock earlier in its
evolution.

Other weaknesses of the simple model include the possibility of
mixing with the high temperature gas that produces X-rays, the
possible effects of thermal conduction, the neglect of dust and the
gradual liberation of refractory elements, the uncertainties in
atomic rates and the approximate treatment of radiative transfer.
Weaknesses in the comparison to observations include the reddening
correction uncertainty, and the scaling for the different apertures
of the different instruments.  We note that a single shock model
could not be expected to match all the line intensities even if the
apertures used in the different observations covered exactly the
same area on the sky, because the echelle spectrum shown by
\citet{danforth01} shows that there are two separate shocks along
the line of sight, redshifted and blueshifted by about 30 \kms.

In spite of all the limitations described above, a simple model of
a 145--155\,\kms\ shock is a reasonable explanation for the
observations, since it matches all the observed lines to within a
factor of 2 (except \iniii, where the discrepancy is slightly higher)
and most to within about 25\%.  Although the match between observations
and model predictions is reasonable, the agreement could be improved
with some tweaking of model parameters, particularly elemental
abundances and cutoff column densities.  However, the parameter
changes would probably lie within the uncertainties.  The fact that
such a simple model comes close to matching the observations may
result from the fact that to first order it is simply conservation
of energy in gas that cools radiatively from the shock temperature.


\subsection{Comparison with other SNRs} 

The set of IR lines detected in the Cusp spectra are useful diagnostics
of SNR shocks.  This is particularly true of the lines from the
singly and doubly ionized species, \neii, \silii, \feii, \neiii,
\siii and \feiii, all of which can be produced in shocks with speeds
as low as 80\,\kms\ \citep{hartigan87}.  We compare the IR line
fluxes from the Cusp region with those measured in other SNRs.  In
Table~\ref{tblsnrs} we have assembled measurements from the published
literature.  Our sample consists of \textit{Infrared Space Observatory}
spectra of IC443 \citep{oliva99a} and RCW 103 \citep{oliva99b}, and
\spitzer\ spectra of W44, W28, IC443C and 3C391 \citep{neufeld07}
and Kes 69, 3C396, Kes 17, G346.6-0.2, G348.5-0.0 and G349.7+0.2
\citep{hewitt09}.

The strongest lines in all the remnants are \silii\,34.80\,\mum\
and \neii\,12.81\mum.  In Table~\ref{tblsnrs} we have arranged the
remnants in order of increasing surface brightness of the sum of
these two lines.  According to this measure, the Cygnus Loop XA
region is the faintest of the sample.  The remaining objects fall
into three groups: at the faint end IC443C and G346.6-0.2 are about
3 times as bright as XA, the brightest remnants, RCW 103 and
G349.7+0.2 are 70 and 180 times as bright as XA, respectively, and
the remaining eight objects are between 7 and 30 times as bright.

The ratio of \neiii\,15.56\,\mum\ to \neii\,12.81\,\mum\ in the XA
region is 1.3, which is the highest value among all the remnants.
The ratio is 1.2 in 3C396, 0.9 in RCW 103, and lies between 0.1 and
0.5 for all other remnants.  Shocks with velocities in the
80--120\,\kms\ range produce \neiii\ to \neii\ ratios between 0.1
and 0.3 when the pre-shock medium is in photoionization equilibrium,
and between 0.5 and 0.9 when it is completely ionized \citep{hartigan87}.
We have explained the higher than predicted ratio in the Cusp
spectrum by invoking an incomplete, $\sim150$\,\kms\ shock.  However,
an alternative explanation is that faster shocks in the vicinity
as well as the hot X-ray emitting plasma provide photoionizing
radiation that increases the \neiii\ to \neii\ flux ratio above the
values expected for only shock excitation.

The \feiii\,22.92\,\mum\ to \feii\,17.93\,\mum\ ratio is another
potential diagnostic of the ionization state, though it can also
depend on the density.  (We do not use the \feii\,25.98\,\mum\ flux
in the comparison since for the other remnants this line is not
well resolved from the \oiv\,25.88\,\mum\ line.) The ratio of the
intensities in the dereddened and scaled Cusp spectrum is 1.60, and
in the uncorrected observed spectrum it is 0.91 (Table~\ref{tblflux})
The \feiii\ line is detected in four other remnants, RCW 103, W44,
W28, and 3C391, and an upper limit is presented for IC443C\@.  The
\feiii\ to \feii\ flux ratio ranges between 0.16 and 0.25, significantly
lower than in the Cusp, and consistent with recombined shocks.

The \neiii\ and \siii\ zones in the post-shock flow are nearly
co-extensive and therefore the relative abundances of Ne and S
influence the flux ratio of \neiii\,15.56\,\mum\ and \siii\,18.71\,\mum.
The ionization potential of S$^{++}$ (34.83\,eV) is lower than that
of Ne$^{+}$ (40.96\,eV) and therefore photoionization can also play
a role in determining the flux ratio of \neiii\ to \siii.  This
trend is detected in the sample of remnants.  For the three objects
with the lowest \neiii\ to \neii\ ratios, i.e. $< 0.16$, the \neiii\
to \siii\ ratios lie between 0.5 and 1.4.  For all the remaining
objects, except one, where the \neiii\ to \neii\ ratio span the
range 0.2--1.3, the \neiii\ to \siii\ ratios lie between 2.4 and
4.4.  The exception is Kes 69, where the ratio is 8.2, indicating
a higher Ne to S abundance in this remnant compared with the others.

The \siv\ line is detected in the XA region and in RCW 103, and in
none of the other remnants.  In RCW 103, the ratio \siv\,10.51\,\mum\
to \siii\,18.71\,\mum\ is about 0.06, which is much lower than for
the XA region, where the ratio is about 0.4.  Overall, we conclude
that the other spectra listed in Table 4 pertain to slower shocks,
than the one at the Cusp, probably because the bright regions chosen
for observation are shocks in denser clouds.


\section{Concluding Remarks}

The Cygnus Loop XA region provides a rich set of shock excited
emission lines across a broad wavelength range.  The analysis of
the morphologically simple Edge shock can be accomplished with
one-dimensional shock models.  The emission line spectrum of the
more complex interior Cusp shock can be reproduced approximately
by a similar simple shock model.  However, the limitations of the
models imply a corresponding limitation in our interpretation of
the shock interaction.

The proximity of the Cygnus Loop allows us to study the IR diagnostics
of a radiative shock in detail.  Our analysis of the XA emission
lines will help in establishing robust diagnostics that can be
applied to other SNRs, where, presumably, the IR spectra are obtained
from regions even more heterogeneous and complex than the Cusp
region.

Our data shows the lack of dust emission in the mid-IR wavelength
range.  Even the slower shocks are presumably effective at destroying
the smaller grains likely to contribute at these wavelengths.  To
probe the dust and possible molecular content of the XA region, we
will need observations at longer wavelengths.  The complex morphology
of the cloud shock interaction will be accompanied by comparably
complicated kinematics.  The identification and use of a kinematic
tracer to map the velocity field in the XA region may be useful in
disentangling the various shock components contributing to the
emission.

By combining the IR spectrum of the Cusp with optical and UV spectra,
we have obtained tight constraints on the shock speed, pre-shock
density, elemental abundances and the column density cutoff, which
corresponds to the age of the shock.  We find that a speed of about
150 \kms\ is needed to match the high ionization lines, rather
efficient destruction of grains is required to match the abundances
of refractory elements, and an age of about 1000 years matches the
column density cut-off and separation between the Cusp and Edge
regions.


\acknowledgments

\begin{figure}
\centering
\includegraphics[height=7.5in]{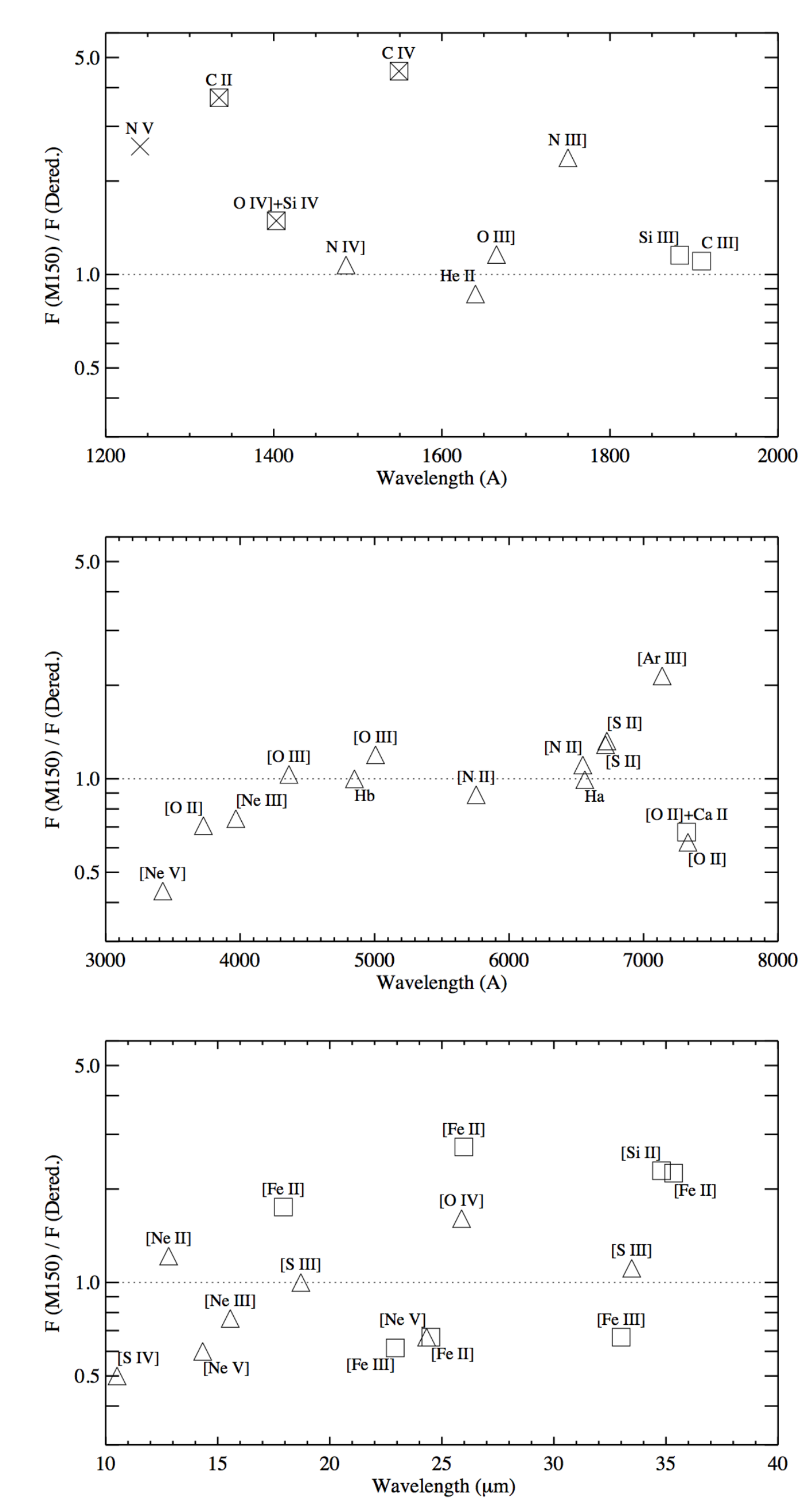}
\caption{The ratio between the model M150 predicted flux and the
observed (scaled, dereddened) value is plotted against wavelength
for the lines listed in Table~\protect\ref{tblflux}.  In the case
where the flux is a sum of two components, the first wavelength
value reported in column 2 of the table is used in the plot.  The
top, middle and bottom plots show the UV, optical and IR lines,
respectively.  Squares indicate lines affected by depletion on to
dust grains, and crosses indicate lines likely to be attenuated by
resonance scattering.  Crosses within squares are used for lines
susceptible to both depletion and resonance scattering.}
\label{fmodobs}
\end{figure}

This work was supported in part by JPL Award 1278412 to the University
of California, Berkeley and North Carolina State University, Raleigh.
We thank the anonymous referee for several useful suggestions, one
of which led to Fig.~\ref{fmodobs}, providing a visual comparison
of model predictions with observations. RS acknowledges support
from USRA at the SOFIA Science Center.  TJG acknowledges support
under NASA contract NAS8-03060 with the Chandra X-ray Center.

Facilities:
\facility{Spitzer Space Telescope.}, \facility{FLWO:1.5m}



\clearpage

\begin{deluxetable}{rcccc}

\tablecaption{Summary of IRS observations \label{tblobs} }
\tablewidth{0pt}
\tablehead{
\colhead{Label\tablenotemark{a}}
  & \colhead{$\alpha_{J2000}$\tablenotemark{b}}
  & \colhead{$\delta_{J2000}$\tablenotemark{b}}
  & \colhead{t$_{exp}$ (SH)\tablenotemark{c}} & \colhead{t$_{exp}$ (LH)\tablenotemark{d}}
}

\startdata
Cusp
     &  $20^{\rm{h}}57^{\rm{m}}14\fs82$ & +31\arcdeg\ 02\arcmin\ 33\farcs4 
     & 481.7\,s & 241.8\,s \\
Edge
     &  $20^{\rm{h}}57^{\rm{m}}25\fs42$ & +31\arcdeg\ 01\arcmin\ 40\farcs5 
     & 481.7\,s & 241.8\,s \\
Background
     &  $20^{\rm{h}}57^{\rm{m}}28\fs50$ & +31\arcdeg\ 02\arcmin\ 04\farcs5 
     & 481.7\,s & 241.8\,s \\
\enddata

\tablenotetext{a} {These labels are used in the text, the images and in Table 2.}
\tablenotetext{b} {The co-ordinates are for the center of the Short-High slit.  The
Long-High slit center is displaced by about 4\arcsec\ in a direction 15\arcdeg\ 
west of north for each position.}
\tablenotetext{c} {Short-High; the slit measures 4.7\arcsec\ $\times$ 11.3\arcsec.}
\tablenotetext{d} {{L}ong-High; the slit measures 11.1\arcsec\ $\times$ 22.3\arcsec.}

\end{deluxetable}


\clearpage

\begin{deluxetable}{lcccccccc}
\tablecolumns{9}
\tablewidth{0pt}
\tablecaption{Cusp Region: Observed and Model Fluxes \label{tblflux} }
\tablehead{
\colhead{Ion}
& \colhead{$\lambda$\tablenotemark{a}}
& \colhead{Obs.\tablenotemark{b}}
& \colhead{Dered.\tablenotemark{c}}
& \colhead{M145\tablenotemark{d}}
& \colhead{M150}
& \colhead{M155a}
& \colhead{M155b}
& \colhead{M155c}
}

\startdata

\sidehead{Ultraviolet Lines}
\pnv\tablenotemark{e}             & 1241       &    85   &   315  &   391  &   814  &  1230  &  1110  &  1050  \\
\pcii\tablenotemark{e}            & 1335       &    33   &   103  &   456  &   382  &   481  &   448  &   433  \\
\ioiv\,+\psiliv\tablenotemark{e}  & 1403       &   162   &   468  &   540  &   697  &   859  &   775  &   736  \\
\iniv            & 1486       &    92   &   248  &   226  &   266  &   280  &   253  &   239  \\
\pciv\tablenotemark{e}            & 1549       &   183   &   476  &  2370  &  2150  &  2040  &  1800  &  1710  \\
\pheii           & 1640       &   115   &   290  &   236  &   251  &   269  &   241  &   229  \\
\ioiii           & 1665       &   145   &   363  &   378  &   421  &   492  &   446  &   425  \\
\iniii           & 1750       &    31   &    77  &   173  &   183  &   195  &   177  &   169  \\
\isiliii         & 1883       &    81   &   209  &   235  &   241  &   297  &   276  &   265  \\
\iciii           & 1909       &   232   &   611  &   701  &   675  &   801  &   731  &   698  \\
\sidehead{Optical Lines}
\nev             & 3425       &    37\tablenotemark{f}   &    62  &    15  &    27  &    44  &    39  &    38  \\
\oii             & 3727       &  1298   &  1432  &   957  &  1012  &  1166  &  1073  &  1009  \\
\neiii           & 3968+3870  &   165   &   176  &   120  &   131  &   152  &   139  &   134  \\
\oiii            & 4363       &    59   &    60  &    56  &    62  &    72  &    65  &    62  \\
H$\beta$         & 4850       &   105   &   100  &   100  &   100  &   100  &   100  &   100  \\
\oiii            & 5007+4959  &   773   &   720  &   771  &   860  &   977  &   883  &   840  \\
\nii             & 5755       &    10   &     9  &     7  &     8  &     9  &     8  &     8  \\
\nii             & 6548+6584  &   429   &   345  &   349  &   382  &   380  &   378  &   379  \\
H$\alpha$        & 6563       &   363   &   292  &   290  &   290  &   290  &   290  &   290  \\
\sii             & 6717       &   153   &   122  &   173  &   157  &   174  &   159  &   144  \\
\sii             & 6727       &   124   &    99  &   139  &   131  &   142  &   153  &   160  \\
\ariii           & 7138       &    18   &    14  &    27  &    30  &    33  &    30  &    29  \\
\oii\,+\pcaii\tablenotemark{g}    & 7320       &    64   &    49  &    34  &    33  &    39  &    41  &    38  \\
\oii             & 7330       &    42   &    32  &    19  &    20  &    24  &    25  &    26  \\
\tablebreak
\sidehead{Infrared Lines: Short-High Module}
\siv             & 10.51      &    9.8  &     8  &     3  &     4  &     4  &     3  &     3  \\
\neii            & 12.81      &   46.9  &    37  &    50  &    45  &    43  &    47  &    49  \\
\nev             & 14.33      &    6.1  &     5  &     2  &     3  &     5  &     5  &     5  \\
\neiii           & 15.56      &   59.5  &    47  &    30  &    36  &    35  &    30  &    29  \\
\feii            & 17.93      &    9.6  &     8  &    14  &    14  &    14  &    16  &    18  \\
\siii            & 18.71      &   24.3  &    19  &    16  &    19  &    14  &    14  &    14  \\
\sidehead{Infrared Lines: Long-High Module}
\feiii           & 22.92      &    8.7  &    13  &     8  &     8  &    11  &     9  &     9  \\
\nev             & 24.32      &    3.9  &     6  &     2  &     4  &     7  &     6  &     6  \\
\feii            & 24.51      &    1.9  &     3  &     2  &     2  &     2  &     3  &     3  \\
\oiv             & 25.88      &   21.7  &    33  &    42  &    53  &    65  &    58  &    55  \\
\feii            & 25.98      &   20.2  &    30  &    82  &    82  &    75  &    69  &    62  \\
\feiii           & 33.00      &    2.2  &     3  &     2  &     2  &     3  &     2  &     2  \\
\siii            & 33.47      &   18.2  &    27  &    25  &    30  &    23  &    19  &    16  \\
\silii           & 34.80      &   61.9  &    93  &   255  &   213  &   207  &   152  &   113  \\
\feii            & 35.34      &    5.1  &     8  &    18  &    18  &    16  &    15  &    14  

\enddata

\tablenotetext{a}{Units: \AA\ for UV and optical lines, \mum\ for IR lines.}
\tablenotetext{b}{Units of $10^{-17}$ \sbu; not scaled between different apertures.}
\tablenotetext{c}{Fluxes dereddened for E$_{B-V}$=0.16,
scaled and normalized (see \S5.2).}
\tablenotetext{d}{See Table \protect\ref{tblshocks} for shock model parameters.}
\tablenotetext{e}{Strongly affected by resonant scattering.
The \psiliv\ contribution to the blend is not included in the model.} 
\tablenotetext{f}{Measured from the narrowband image of \cite{szentgyorgyi00}; scaled, normalized flux = 39 for
E$_{B-V}=0.08$.}
\tablenotetext{g}{The \pcaii\ line contribution is included in the model.}

\end{deluxetable}


\clearpage

\begin{deluxetable}{lccccc}
\tablecolumns{6}
\tablewidth{0pt}
\tablecaption{Shock Model Parameters \label{tblshocks} }
\tablehead{
\colhead{}
& \colhead{M145}
& \colhead{M150}
& \colhead{M155a}
& \colhead{M155b}
& \colhead{M155c}
}

\startdata

\sidehead{Input Parameters}
$v_{shock}$(\kms)    & 145 & 150 & 155 & 155 & 155 \\
n$_{0}$(cm$^{-3}$)   & 2.5 & 5.0 & 2.5 & 5.0 & 7.9 \\
B$_{0}$($\mu$G)      & 1.0 & 3.0 & 1.0 & 1.0 & 1.0 \\
\sidehead{Value at Cutoff}
log N(H)/(cm$^{-2}$) & 18.0 & 18.1 & 18.0 & 18.0 & 18.0 \\
T(K)                 & 7300 & 7700 & 7400 & 7300 & 7500

\enddata

\tablecomments{H ionized, He singly ionized in the
pre-shock gas for all models.}

\end{deluxetable}


\clearpage

\begin{deluxetable}{rccccc}

\tablecaption{Selected IR flux and flux ratios of SNRs \label{tblsnrs} }
\tablewidth{0pt}
\tablehead{
\colhead{SNR} 
& \colhead{\neii+\silii\tablenotemark{a}} 
& \colhead{\neiii/\neii}
& \colhead{\neiii/\siii} 
& \colhead{Ref.}
}

\startdata

CygLoop     &   110   &    1.27   &   2.4  &  1 \\
IC443C\tablenotemark{b}   &   270   &    0.27   &   3.0  &  4 \\
G346.6-0.2  &   290   &    0.07   &   1.4  &  5 \\
W44         &   720   &    0.11   &   0.5  &  4 \\
Kes 17      &  1090   &    0.35   &   4.4  &  5 \\
3C396       &  1110   &    1.24   &   3.5  &  5 \\
Kes 69      &  1150   &    0.24   &   8.2  &  5 \\
IC443\tablenotemark{b}    &  1260   &    0.56   &   2.5  &  2  \\
W28         &  1560   &    0.16   &   1.0  &  4 \\
G348.5-0.0  &  3120   &    0.20   &   4.2  &  5 \\
3C391       &  3230   &    0.38   &   2.6  &  4 \\
RCW 103      &  7420   &    0.89   &   3.1  & 3 \\
G349.7+0.2  & 19620   &    0.34   &   3.8  &  5

\enddata

\tablecomments{The lines reported are \neii\,12.81\,\mum,
\neiii\,15.56\,\mum, \siii\,18.71\,\mum\ and \silii\,34.80\,\mum.}
\tablerefs{1. this work, 2. Oliva et al.\ 1999a, 3. Oliva et al.\
1999b, 4. Neufeld et al.\ 2007, 5. Hewitt et al.\ 2009}
\tablenotetext{a}{Units of $10^{-17}$ \sbu} \tablenotetext{b}{The
region IC443C is distinct from IC443, though both are part of the
same remnant.}

\end{deluxetable}


\clearpage

\newpage

\newpage

\newpage

\newpage

\newpage

\newpage

\newpage

\newpage


\end{document}